\documentstyle[tighten,preprint,aps,amsfonts]{revtex}
\DeclareMathSymbol{\shortparallel}{\mathrel}{AMSb}{"71}
\DeclareMathSymbol{\restriction}{\mathrel}{AMSa}{"16}

\newcommand{\ee}{\end{document}}
\newcommand{\llabel}[1]{\label{#1}}
\begin{document}
\draft
\title{The Scalar Curvature of the Bures Metric on 
the Space of Density Matrices}
\author{J.~Dittmann\thanks{
e-mail correspondence: {\tt dittmann@mathematik.uni-leipzig.de}}}
\address{Mathematisches Institut, Universit\"at Leipzig\\
Augustusplatz 10/11, 04109 Leipzig,
Germany}
\maketitle
\begin{abstract}
The Riemannian Bures metric on the space of (normalized) complex positive 
matrices is used for parameter estimation of mixed quantum states 
based on repeated measurements
just as the Fisher information in classical statistics. It appears 
also in the concept of purifications of mixed states in quantum physics. 
Therefore, and also for mathematical reasons,
it is natural to ask for curvature properties of this Riemannian metric.
Here we determine  its  scalar curvature and  Ricci tensor and prove a 
lower bound 
for the curvature on the submanifold of trace one matrices. This bound  
is achieved for the 
maximally mixed state, a further hint for the 
statistical meaning of the scalar curvature.  
\end{abstract}
\section{Introduction}
Let $\cal D$ denote the space of complex positive 
$n{\times n}$-matrices for a fixed $n$ 
and ${\cal D}^1$ the submanifold of trace one matrices. ${\cal D}^1$ 
represents the  space of nondegenerate mixed states of a $n$-dimensional 
quantum system. The tangent space at $\varrho\in {\cal D}$ 
(resp.~${\cal D}^1$)
consists of all Hermitian (traceless) matrices. These manifolds carry 
the so called  Riemannian Bures metric $g$ defined by
$$
g_\varrho(X,Y)=\frac{1}{2}\,{\rm Tr}\,X G\,,
\qquad X,Y\in{\rm T}_\varrho{\cal D}\,,
$$
where $G$ is the  (unique, by the Sylvester-Rosenblum theorem, see 
\cite{BR97}) solution of $\varrho G+G\varrho=Y$. 
It should be mentioned, 
that $g$ is also
well defined on manifolds of all $\varrho\geq 0$ of fixed  rank, but we 
will deal only with the maximal rank.
This Riemannian metric  was introduced by Uhlmann in generalizing the Berry 
phase to mixed states, \cite{UH86,Uh91,Uh92b}. He was led to this metric by 
asking for curves on minimal length purifying  a given path of 
densities. Later on this metric 
appeared also in other contexts, see e.~g.~\cite{Braunstein,Petz}.

The restriction of $g$ to the manifold of trace one diagonal matrices, 
i.~e.~to the manifold of all probability distributions on a $n$-point set, 
is (up to the  factor 1/4)
just the Fisher metric known from classical statistics, 
see e.~g.~\cite{Amari,Friedrich}. 
Similarly to this case the Bures metric is related to the 
statistical distance of 
quantum states, see \cite{Braunstein,Braunstein96}. 
Roughly speaking, both metrics give a lower bound for the variance of an 
optimal parameter estimator. Thus the Bures metric 
generalizes the classical Fisher information to the quantum case. Among 
other generalization, namely the so called monotone metrics 
(i.~e.~metrics decreasing under stochastic mappings), 
\cite{Petz}, the Bures metric is minimal, and  
it seems to play a distinguished role also for other reasons, see 
\cite{Di98,UD}. Partial results concerning the curvature of 
the Bogoliubov metric,
another monotone metric, were obtained in \cite{Petz96}.

Several authors, e.~g.~\cite{Twamley,Paraoanu}, suggested 
that the  scalar curvature has a quantum  statistical meaning 
as a  measure of local distinguishability of states in the sense, that 
regions of small curvature require many measurements for distinguishing 
between neighboring states. But this is still in progress and, up to now, 
no statistical equation or estimation involving the scalar curvature seems 
to be available. 
However, we show that the scalar curvature is minimal for the 
maximally mixed state  $\frac{1}{n}\bbox{1}$ and that it diverges nearby 
pure states, further hints for the suggested statistical 
meaning. 

We determine here the Ricci tensor and  the scalar curvature  
(Propositions 2 and 3) completing the list of basic 
local curvature quantities of the Bures metric. 

\noindent{\bf Notations:} 
The eigenvalues of a positive matrix $\varrho$ are denoted by $\lambda_i$.
Thus, if we assume $\varrho$ to be  diagonal, then 
$\varrho=diag(\lambda_1,\dots,\lambda_n)$. 
Bold letters are used for operators  acting on 
matrices. They will   depend on $\varrho$, so that they 
actually represent fields of operators called by several authors 
superoperators. However,   we frequently  suppress this dependence for 
brevity of  notation similarly to vector field and other  
quantities. In particular, ${\bf L}_\varrho$ and  ${\bf R}_\varrho$ denote 
the operators of left and right multiplication by $\varrho$ and 
$\frac{1}{{\bf L}+{\bf R}}$ is the inverse operator of ${\bf L}+{\bf R}$ 
(denoted by ${\cal R}^{-1}_{\hat{\rho}}$ in \cite{Braunstein}). 
This operator appears in many of the following formulae
and is a serious obstruction for using coordinates in handling the 
Bures metric which now reads 
$$g=
\frac{1}{2}\,{\rm Tr}\,{\rm d }\varrho\frac{1}{\bf L+R}({\rm d }\varrho)\,.
$$
However, from the theory of matrix equations, 
\cite{Smith,BR97}, some explicit formulae  for this 
operator and the metric can be derived, \cite{Di98a}.

\section{The Riemannian Curvature Tensor}
In this section we explain some results 
concerning  the 
Riemannian curvature tensor of the Bures metric and introduce on 
this occasion some further notations. 
A brief communication 
of this results appeared in \cite{Di93}. Proofs and more details
can be found  in \cite{Di93a}.  

The manifold $\cal D$ is an open subset of the  space of Hermitian 
matrices and all tangent spaces ${\rm T}_\varrho{\cal D}$ are
identified with this real vector space. Thus we regard
vector fields $X,Y,\dots $ on ${\cal D}$  as 
functions $\varrho\mapsto X_\varrho$ on ${\cal D}$ with 
Hermitian   (traceless for ${\cal D}^1$) values. 
The  flat covariant derivative $\nabla^f$ on $\cal D$ 
inherited from the affine structure of the Hermitian matrices  is simply 
the derivation along  straight lines; 
$(\nabla^f_XY)_\varrho=
\lim_{t\to0}(Y_{\varrho+tX_\varrho}-Y_\varrho)/t
$. In particular, $\nabla^f_XN=X$, where $N$ is the vector field 
defined by $N_\varrho:=\varrho$. It is perpendicular w.~r.~to the Bures 
metric to the submanifold ${\cal D}^1$. This 
allows for determining curvature quantities of ${\cal D}^1$ from that of 
$\cal D$ by  the Gauss equation. Quantities 
with superscript 1 will always refer to ${\cal D}^1$ .

We denote by  $\nabla$ resp.~$\nabla^1$ the covariant derivative of the 
Levi-Civita connection of the Bures metric on ${\cal D}$ 
resp.~${\cal D}^1$. In \cite{Di93a} it was shown that
\begin{mathletters}\label{nabla}
\begin{eqnarray}
\nabla_XY&=&\llabel{nablaa}\nabla^f_XY-
{\textstyle\frac{1}{\bf L+R}}(X)N{\textstyle\frac{1}{\bf L+R}}(Y)-
{\textstyle\frac{1}{\bf L+R}}(Y)N{\textstyle\frac{1}{\bf L+R}}(X)\\
\nabla^1_XY&=&\llabel{nablab}\nabla_XY+2g(X,Y)N\;,
\end{eqnarray}
\end{mathletters}
where on the right hand side  of (\ref{nablaa}) appears the usual 
product of matrix valued functions.
Of course, in order to apply (\ref{nablab}) one  must   extend the 
vector fields $X$ and $Y$ on ${\cal D}^1$ to a neighborhood of 
${\cal D}^1$, but the result will not depend on this extension.  
(\ref{nablaa}) corresponds to the well known equation
$$
\nabla_{\partial_i}a^j \partial_j=
{\partial_i}(a^j){\partial_j}+a^j\Gamma_{ij}^k{\partial_k},
$$
which relates the covariant derivative to the flat 
derivative induced by a local parametrization.
The calculations of the next section are based on the following  
Proposition derived from (\ref{nabla}). 

\medskip\noindent
{\bf Proposition 1 :}\newline
{\it The curvature   tensor field of the Bures metric  on 
${\cal D}$ resp.~ ${\cal D}^1$ is given by} 
\begin{mathletters}\llabel{curv}
\begin{eqnarray}
{\cal R}(W,Z,X,Y)&:=&
g\left(\nabla_X\nabla_Y-\nabla_Y\nabla_X-\nabla_{[X,Y]},W\right)
\nonumber\\
&=&\;\,2\,g\left({\rm i\,}{\bf LR}
\left[{\textstyle\frac{1}{\bf L+R}}X,{\textstyle\frac{1}{\bf L+R}}Y\right]
,{\rm i}
\left[{\textstyle\frac{1}{\bf L+R}}W,{\textstyle\frac{1}{\bf L+R}}Z\right]
\right)
\nonumber\\
&&{}+
g\left({\rm i\,}{\bf LR}
\left[{\textstyle\frac{1}{\bf L+R}}Z,{\textstyle\frac{1}{\bf L+R}}Y\right]
,{\rm i}
\left[{\textstyle\frac{1}{\bf L+R}}W,{\textstyle\frac{1}{\bf L+R}}X\right]
\right)\nonumber\\
&&{}-
g\left({\rm i\,}{\bf LR}
\left[{\textstyle\frac{1}{\bf L+R}}Z,{\textstyle\frac{1}{\bf L+R}}X\right]
,{\rm i}
\left[{\textstyle\frac{1}{\bf L+R}}W,{\textstyle\frac{1}{\bf L+R}}Y\right]
\right)\,,\llabel{curv1}\\
&&\nonumber\\
{\cal R}^1(W,Z,X,Y)=&{\cal R}&(W,Z,X,Y)
+g(Y,Z)g(X,W)-g(X,Z)g(Y,W)\,.
\end{eqnarray}
\end{mathletters}
\hfill$\Box$

\noindent
Note the different meaning of commutators in the  equations above. 
In (\ref{curv1}) it is  pointwise the usual matrix commutator,
$[X,Y]_\varrho:=X_\varrho Y_\varrho-Y_\varrho X_\varrho$. All further 
commutators will be understood in this sense. An immediate consequence of 
Proposition 1 is 

\medskip
\noindent{\bf Corollary 1:}
{\newline\it Let $p$ be the plane generated by two tangent vectors 
$X$ and $Y$. Then the sectional curvature is given by}
\begin{mathletters}
\begin{eqnarray*} 
{\cal K}(p)&:=&
\frac{{\cal R}(X,Y,X,Y)}{g(X,X)g(Y,Y)-g(X,Y)^2}\nonumber\\
&&\nonumber\\
&=&\frac{
3g\left({\rm i\,}{\bf LR}
\left[{\textstyle\frac{1}{\bf L+R}}X,{\textstyle\frac{1}{\bf L+R}}Y\right]
,{\rm i}
\left[{\textstyle\frac{1}{\bf L+R}}W,{\textstyle\frac{1}{\bf L+R}}Z\right]
\right)}{g(X,X)g(Y,Y)-g(X,Y)^2}\\
&&\nonumber\\
{\cal K}^1(p)&=&K(p)+1 \,.
\end{eqnarray*}
\end{mathletters}
\hfill$\Box$\newline
Finally we mention that for $n=2$ the Riemannian manifold
$({\cal D}^1,g)$ is isometric to an open 
half 3-sphere of radius 1/2, \cite{Uh92b}. 
The  geometry for $n>2$ is much  more
complicated, e.~g.~${\cal D}^1$ is not locally symmetric, \cite{Di93a}.

\section{Ricci Tensor and Scalar curvature}
In order to determine the Ricci tensor and the scalar curvature we have to 
calculate traces of the curvature given by Proposition 1. 
For clarity we will
distinguish in the notation between the trace of matrices and the trace of 
operators acting on matrices. We will treat 
simultaneously the normalized and the unnormalized case. 
For brevity we include 
in brackets additional terms corresponding to the normalized case . 

First we determine 
 the curvature mapping, 
 also denoted by $\cal R$, which is given by $g({\cal R}(X,Y)Z,W)={\cal 
R}(W,Z,X,Y)$. For this purpose we have to separate $W$ in (\ref{curv1}) as a 
single argument of $g$. Using the  definition of $g$ and the selfadjointness 
of $\bf L $ and $\bf R$ w.~r.~to the Hilbert-Schmidt product we obtain 
after a straightforward calculation 
\begin{eqnarray}
{\cal R}^{(1)}(X,Y)Z&=&\,2
\left[{\textstyle\frac{1}{\bf L+R}}Z,{\textstyle\bf\frac{LR}{L+R}}
\left[{\textstyle\frac{1}{\bf L+R}}Y,{\textstyle\frac{1}{\bf L+R}}X\right]
\right]\nonumber\\
&&+
\left[{\textstyle\frac{1}{\bf L+R}}X,{\textstyle\bf\frac{LR}{L+R}}
\left[{\textstyle\frac{1}{\bf L+R}}Y,{\textstyle\frac{1}{\bf L+R}}Z\right]
\right]\nonumber\\
&&\llabel{curvmap}
+\left[{\textstyle\frac{1}{\bf L+R}}Y,{\textstyle\bf\frac{LR}{L+R}}
\left[{\textstyle\frac{1}{\bf L+R}}Z,{\textstyle\frac{1}{\bf L+R}}X\right]
\right]\quad+\Big(\; g(Y,Z)X-g(X,Z)Y\;\Big)\,.
\end{eqnarray}
The Ricci tensor is defined by
$$
{\it Ricci}(Y,Z):={\bf Tr}\,\{\,X\mapsto{\cal R}(X,Y)Z\,\}\,. $$
Eliminating $X$ in (\ref{curvmap}) yields 
\begin{eqnarray}
{\it Ricci}^{(1)}(Y,Z)&=&
{\bf Tr}\,\Big\{\;
2\,{\bf ad}{\textstyle\frac{1}{\bf L+R}}Z\circ
{\textstyle\bf\frac{LR}{L+R}}\circ
{\bf ad}{\textstyle\frac{1}{L+R}}Y\circ
{\textstyle\frac{1}{\bf L+R}}\nonumber\\
&&\qquad +\,{\bf ad}{\textstyle\bf\frac{LR}{L+R}}\!
\left[{\textstyle\frac{1}{\bf L+R}}Z,{\textstyle\frac{1}{\bf L+R}}Y\right]
\circ{\textstyle\frac{1}{\bf L+R}}\nonumber\\
&&\llabel{Ricci}
\qquad +\,{\bf ad}{\textstyle\frac{1}{\bf L+R}}Y\circ
{\textstyle\bf\frac{LR}{L+R}}\circ
{\bf ad}{\textstyle\frac{1}{\bf L+R}}Z\circ
{\textstyle\frac{1}{\bf L+R}}
\;\Big\}
\quad+\Big(\; (n^2-2)g(Y,Z)\;\Big).
\end{eqnarray}
This equation requires  some comments. ${\bf ad}V$ 
denotes the usual commutation operator, ${\bf ad}V(W):=[V,W]$, and we have 
to do with compositions of operators. 
The trace should be regarded, originally, 
on the real tangent spaces, that means on the Hermitian matrices, 
traceless or not. But the normal direction generated by $\varrho$
does not give any contribution to the trace in $Ricci(Y,Z)$ because 
${\cal R}(X,Y)Z$ vanishes for $X_\varrho:=\varrho$. The additional term in 
the normalized case is the trace of 
$X\mapsto g(Y,Z)X-g(X,Z)Y$ on the $(n^2{-}1)$-dimensional space of traceless 
Hermitian matrices. Finally, the trace of a real operator equals the 
trace of its complexification. Therefore we can take the trace in 
(\ref{Ricci})  on  
all complex $n{\times}n$-matrices.

To continue the determination of the Ricci tensor we notice that the 
second
term of the trace in (\ref{Ricci}) vanishes, 
because  ${\bf Tr\,ad}V{\circ} ({\bf L+R})^{-1}=0$ for 
all $V$. 
Indeed, we can suppose that $\varrho$ is diagonal. Then
$$
{\bf Tr\,ad}V\circ{\textstyle\frac{1}{\bf L+R}}=
\sum_{i,j}\,\langle{\rm E}_{ij},
[V,{\textstyle\frac{1}{\bf L+R}}{\rm E}_{ij}]
\rangle=
\sum_{i,j}\,\frac{1}{\lambda_i+\lambda_j}\langle
{\rm E}_{ii}-{\rm E}_{jj},V\rangle=0\,.
$$
The remaining expression in (\ref{Ricci}) must be symmetric in $Y$ and $Z$, 
since the Ricci tensor is symmetric. Hence (\ref{Ricci}) reduces to
\begin{equation}\llabel{Ricci2}
{\it Ricci}^{(1)}(Y,Z)=
3\,{\bf Tr}\;
{\bf ad}{\textstyle\frac{1}{\bf L+R}}Y\circ
{\textstyle\bf\frac{LR}{L+R}}\circ
{\bf ad}{\textstyle\frac{1}{\bf L+R}}Z\circ
{\textstyle\frac{1}{\bf L+R}}
\quad+\Big(\; (n^2-2)g(Y,Z)\;\Big).
\end{equation}

The Ricci tensor can be represented as
$Ricci(Y,Z)=g\left(Y,{\bf F}_{Ricci}(Z)\right)\,,$
where the Ricci mapping ${\bf F}_{Ricci}$ is a  field of operators 
self-adjoint w.~r.~to the Bures metric and whose trace is the scalar 
curvature. We cannot expect that ${\bf F}_{Ricci}$ is a simple expression 
in terms of $\bf L$ and $\bf R$, e.~g.~like 
${\bf LR(L+R)}^{-1}$. Indeed, if $\varrho$ is diagonal  we obtain 
from (\ref{Ricci2}) using the standard basis after a 
simple calculation
\begin{mathletters}\llabel{zw}
\begin{equation}\llabel{Ricci1}
Ricci(Y,Z)=
3\sum_{i,j,k}\frac{Y_{ji}\lambda_kZ_{ij}}{(\lambda_i+\lambda_j)
(\lambda_i+\lambda_k)(\lambda_k+\lambda_j)}-
\frac{3}{2}\sum_{i,j}\frac{Y_{ii}Z_{jj}}{(\lambda_i+\lambda_j)^2}
\end{equation}
and 
\begin{equation}
{\bf F}_{Ricci}(Z)=
6\sum_{i,j,k}\frac{\lambda_k}{(\lambda_i+\lambda_k)(\lambda_k+\lambda_j)}
Z_{ij}{\rm E}_{ij}-
6\sum_{i,j}\frac{\lambda_i}{(\lambda_i+\lambda_j)^2}
Z_{jj}{\rm E}_{ii}
\end{equation}
\end{mathletters}
for $Y,Z\in{\rm T}_\varrho{\cal D}$.
To express the Ricci mapping for a general $\varrho$ we need the 
following natural   mappings:
$$
{\frak m},{\frak m}_{\rm o}:{\cal A}\otimes {\cal A}
\longrightarrow{\cal A}\,,\qquad
\Delta:{\cal A}\longrightarrow {\cal A}\otimes {\cal A}
\,,\qquad
{\cal A}:={\rm M}_{n{\times n}}({\Bbb C})\,,
$$
where
$\frak m$ is the usual multiplication, 
${\frak m}_{\rm o}$ the opposite  multiplication,
${\frak m}_{\rm o}(X\otimes Y)=YX$, and $\Delta$ the 
comultiplication.
It is  the dual 
of $\frak m$ if we identify $\cal A$ and $\cal A^*$ via
$A\mapsto\langle A,\cdot\,\rangle$. 
Explicitly,
$$\Delta({\rm E}_{ij})=\sum_k{\rm E}_{ik}\otimes{\rm E}_{kj}\,.$$
It is obvious that these mappings are equivariant w.r.~to the adjoint 
action of the unitary group, 
e.~g.~$\Delta(uXu^*)=({\bf Ad}u\otimes{\bf Ad}u)\,\Delta(X)$.
Using these mappings we have:

{\pagebreak\samepage \noindent
{\bf Proposition 2:}
\begin{mathletters}\label{PropRicci}
\begin{equation}\llabel{Ricci4}
Ricci^{(1)}(Y,Z)=g\left(Y,{\bf F}_{Ricci}^{(1)}(Z)\right)\,,
\end{equation}
where
\begin{equation}\llabel{Ricci3}
{\bf F}_{Ricci}^{(1)}=
6({\frak m}-{\frak m}_{\rm o})\circ
\left(
{\textstyle\frac{\bf LR}{\bf L+R}}\otimes{\textstyle\frac{1}{\bf L+R}}+
{\textstyle\frac{1}{\bf L+R}}\otimes{\textstyle\frac{\bf LR}{\bf L+R}}
\right)\circ \Delta\circ
{\textstyle\frac{1}{\bf L+R}}
\quad+\Big(\; (n^2-2)\,{\bf Id}\;\Big).
\end{equation}
\end{mathletters}}
Proof: We prove the unnormalized case, the additional term in 
the normalized one is clear  from (\ref{Ricci}). 
If $\varrho$ is diagonal  the last equation follows by comparing 
(\ref{PropRicci}) with (\ref{Ricci1}).
For general $\varrho$ it is sufficient to remark 
that the right hand side of (\ref{Ricci3}) is a (1,1)-tensor field 
on $\cal D$ invariant under the U$(n)$-conjugation. This implies 
the invariance of the right hand side of (\ref{Ricci4}).
\hfill$\Box$

\smallskip
Now we proceed with the scalar curvature ${\cal S}={\bf Tr\, F}_{Ricci}$.
Again, the normal direction does not give a contribution to the trace 
and we can take it on all complex matrices. We will use some obvious 
algebraic relations between the multiplication operators, e.~g.~ 
\begin{eqnarray*}
{\frak m}\circ({\bf L\otimes Id})
&=&
{\bf L}\circ{\frak m}\circ
({\bf Id\otimes Id})\,,\\
{\frak m}\circ({\bf R\otimes Id})
&=&
{\frak m}\circ({\bf Id\otimes L})\,,\\
{\frak m_{\rm o}}\circ
({\bf R\otimes Id})
&=&
{\bf R}\circ{\frak m}_{\rm o}\circ
({\bf Id\otimes Id})\,,\\
{\frak m_{\rm o}}\circ({\bf L\otimes Id})
&=&
{\frak m_{\rm o}}\circ({\bf Id\otimes R})\,,
\end{eqnarray*}
and obtain from (\ref{Ricci3})
\begin{eqnarray}
{\cal S}&=&{\bf Tr\, F}_{Ricci}\nonumber\\&=&
6\,{\bf Tr}\,({\bf L+R)}\circ
\Big\{
{\frak m}\circ({\bf R {\otimes} Id})-
{\frak m}_{\rm o}\circ({\bf Id {\otimes} R})
\Big\}\circ
\left(
{\textstyle\frac{1}{\bf L+R}}{\otimes}{\textstyle\frac{1}{\bf L+R}}
\right)\circ \Delta\circ
{\textstyle\frac{1}{\bf L+R}}\nonumber\\
&=&\llabel{Trace}
6\,{\bf Tr}\,
\Big\{
{\frak m}\circ({\bf R {\otimes} Id})-
{\frak m}_{\rm o}\circ({\bf Id {\otimes} R})
\Big\}\circ
\left(
{\textstyle\frac{1}{\bf L+R}}{\otimes}{\textstyle\frac{1}{\bf L+R}}
\right)\circ \Delta\,.
\end{eqnarray}
The evaluation of this trace yields:

\medskip\noindent
{\bf Proposition 3:} {\it The scalar curvature on ${\cal D}$
resp.~${\cal D}^{1}$
equals
\begin{mathletters}\llabel{Prop3}
\begin{eqnarray}
{\cal S}^{(1)}_\varrho&=&
6\,{\rm 
Tr}\,\varrho\,\frac{\chi'_\varrho(-\varrho)^2}{\chi_\varrho(-\varrho)^{2}}-
\frac{3}{2}\,{\rm Tr}\;\varrho^{-1}\quad+
\Big(\; (n^2-1)(n^2-2)\;\Big)\llabel{Prop3a}\\
&=&{\rm Tr}\,h_\varrho(\varrho)
\quad+\Big(\; (n^2-1)(n^2-2)\;\Big)\,,\llabel{Prop3b}
\end{eqnarray}
\end{mathletters}
where $\chi_\varrho$ is the characteristic polynomial of $\varrho$, 
$\chi'_\varrho$ its derivative and $h_\varrho$ the function  given by
$$h_\varrho(t):=6\,t\Big({\rm 
Tr}\,\frac{1}{\varrho+t\bbox{1}}\Big)^2-\frac{3}{2t}\,.
$$ 
}

\noindent
{\bf Remark:} {\it $\chi_\varrho(-\varrho)$ is, in fact, invertible since 
$\chi_\varrho(-t)=\prod(\lambda_i+t)$ implies 
$\chi_\varrho(-\lambda_j)>0$ for all eigenvalues.}\hfill$\Box$

\medskip\noindent
Proof: It is sufficient to prove the assertion for  diagonal $\varrho$.
For such $\varrho$ it is easy to calculate the trace (\ref{Trace}) and 
we obtain
\begin{eqnarray}
{\cal S}_\varrho&=&6\,\sum_{i,j,k}
\frac{\lambda_k}{(\lambda_i+\lambda_k)(\lambda_k+\lambda_j)}-
\frac{3}{2}\,\sum_i \frac{1}{\lambda_i}
=
6\,\sum_k\lambda_k
\Big(\sum_i
\frac{1}{\lambda_i+\lambda_k}\Big)^2
-\frac{3}{2}\,{\rm Tr}\,\varrho^{-1}\llabel{proof}\,.
\end{eqnarray}
This is in accordance with formulae (\ref{Prop3}).
The additional term in the normalized case is obvious by 
(\ref{Ricci3}).
\hfill$\Box$

The scalar curvature   depends only on the invariants of 
$\varrho$. In order to express $\cal$ it in terms of  
invariants we introduce the following  matrix depending on $\varrho$:
$$
{\cal E}:=\left[{\cal E}_{ij}\right]_{i,j=1}^n\,,\qquad
{\cal E}_{ij}:=
\left\{
\begin{array}{lll}
1&\quad&\mbox{for }i+1=j\\
(-1)^{n-j}e_{n+1-j}&&\mbox{for }i=n\\
0&&\mbox{otherwise}
\end{array}
\right.
$$
where $e_i$ is the elementary invariant  of degree $i$ of $\varrho$,
i.~e.~$\chi(t)=\sum_{i=0}^ne_{n-i}(-t)^i$. 
Since  $\cal E_\varrho$ has the 
same characteristic polynomial as $\varrho$  both matrices 
are conjugate provided the eigenvalues of $\varrho$ are different. 
Thus, at least for such points, we get from Proposition 3

\medskip\noindent
{\bf Corollary 2:} 
\begin{eqnarray*}
{\cal S}^{(1)}&=&
6\,{\rm 
Tr}\,{\cal E}\,\frac{\chi'(-{\cal E})^2}{\chi(-{\cal E})^{2}}-
\frac{3}{2}\,{\rm Tr}\;{\cal E}^{-1}\quad+
\Big(\; (n^2-1)(n^2-2)\;\Big)\,,\\
&=&{\rm Tr}\,h_{\cal E}({\cal E})
\quad+\Big(\; (n^2-1)(n^2-2)\;\Big)\,,\quad \mbox{where}\\
h_{\cal E}(t)&:=&6\,t\Big({\rm 
Tr}\,\frac{1}{{\cal E}+t\bbox{1}}\Big)^2-\frac{3}{2t}\,.\nonumber
\end{eqnarray*}
Since the set of $\varrho$ with different eigenvalues is dense,
the Corollary is true for all points by  continuity of the curvature.

A further consequence of Proposition 3 is the following lower bound 
for the scalar curvature in the normalized case:

\medskip\noindent
{\bf Corollary 3:} {\it 
$$
{\cal S}^1_\varrho \geq \frac{(5n^2-4)(n^2-1)}{2}\,.
$$
For $n>3$ equality holds iff $\varrho=\frac{1}{n}\bbox{1}$. For
$n=2$ the scalar curvature equals 24 for all $\varrho$. }

\bigskip\noindent Proof:  
The eigenvalues of $\varrho\in{\cal D}^1$ satisfy $\sum\lambda_i=1$
and we have
\begin{eqnarray}
&&\sum_k\lambda_k
\Big(\sum_i
\frac{1}{\lambda_i+\lambda_k}\Big)^2-
\frac{1}{4}\,\sum_k\frac{1}{\lambda_k}=
\sum_k\lambda_k
\Big(\sum_{i\atop i\neq k}
\frac{1}{\lambda_i+\lambda_k}\Big)^2+
\sum_{i\neq k}
\frac{1}{\lambda_i+\lambda_k}\nonumber\\
&\geq&
\Big(\sum_{i\neq k}
\frac{\lambda_k}{\lambda_i+\lambda_k}\Big)^2+
\sum_{i\neq k}
\frac{1}{\lambda_i+\lambda_k}\geq
\frac{n^2(n-1)^2}{4}+\frac{n^2(n-1)}{2}=
\frac{n^2(n^2-1)}{4}\,.\nonumber
\end{eqnarray}
Here we used the Schwartz inequality, the relation
$$
\sum_{i\neq k}
\frac{\lambda_k}{\lambda_i+\lambda_k}=
\sum_{i,k}
\frac{\lambda_k}{\lambda_i+\lambda_k}-\frac{n}{2}=
\frac{n^2}{2}-\frac{n}{2}=
\frac{n(n-1)}{2}$$ and  
the fact that the arithmetic mean
of all $1/(\lambda_i+\lambda_k)$, $i\neq k$, is greater than or equal 
to the harmonic mean which equals $n/2$. Hence, equations 
(\ref{Prop3}) and (\ref{proof}) imply 
$${\cal S}^1_\varrho\geq \frac{3}{2}\,n^2(n^2-1)+(n^2-1)(n^2-2)=
\frac{(5n^2-4)(n^2-1)}{2}\,.$$
Moreover, the bound   is achieved for $\varrho=\frac{1}{n}\bbox{1}$.
Finally we note that for $n=2$ the above estimations are, 
in fact, equations (${\cal S}^1=24$). For higher $n$
this can hold only iff all  $\lambda_i+\lambda_k$, $i\neq k$, 
are equal, i.~e.~iff $\lambda_i=1/n$. Hence,
$\varrho=\frac{1}{n}\bbox{1}$ is the only minimal point.\hfill$\Box$

There is no upper bound for $n>2$. Indeed, by (\ref{proof}) the scalar 
curvature equals up to a constant the sum of all 
$6\lambda_k/((\lambda_i+\lambda_k)(\lambda_k+\lambda_j))$,
where   not all indices are equal. Therefore, 
${\cal S}^1$ tends to infinity iff $e_{n-1}$ tends to zero,
because
$e_{n-1}$ is the sum of all $\lambda_{i_1}\dots\lambda_{i_{n-1}}$,
$i_1<i_2<\dots<i_{n-1}$. Roughly speaking ${\cal S}^1$ diverges if we get 
close to   density matrices of rank  $k<n-1$, 
in particular, if we get close to a pure state.

\bigskip\noindent
{\bf Example: }We consider the scalar curvature on ${\cal D}^1$  for 
$n=3$
using Corollary 2: 
We have to set $e_1=1$. Then 
{\renewcommand{\arraystretch}{0.7}
$$\chi(t)=-t^3+t^2-e_2t+e_3\,, \qquad
{\cal E}
=\left(\begin{array}{ccc}
0&1&0\\
0&0&1\\
e_3&-e_2&1
\end{array}\right)\,,
$$
$$
\chi(-{\cal E})=
2\left(\begin{array}{ccccc}
e_3&&0&&1\\
e_3&&e_3-e_2&&1\\
e_3&&e_3-e_2&&1+e_3-e_2
\end{array}\right)\,,\qquad
\chi'(-{\cal E})=
\left(\begin{array}{ccccc}
-e_2&&-2&&-3\\
-3e_3&&2e_2&&-5\\
-5e_3&&5e_2-3e_3&&2e_2-5
\end{array}\right)\,,
$$
and we obtain 
$$
{\cal S}^1=
6\,{\rm Tr}\,{\cal E}\,\chi'(-{\cal E})^2\chi(-{\cal E})^{-2}
-\frac{3}{2}\,{\rm Tr}\;{\cal E}^{-1}+56=
2\:\frac{28\,e_3-49\,e_2-9}{e_3-e_2}\,.
$$}

Similarly we get for $n=4$:
$$
{\cal S}^1=
6\:\frac{63\,e_4+35\,e_3^2-43\,e_2e_3-7\,e_3-3\,e_2^2}{e_4+e_3^2-e_2e_3}\,.
$$
\acknowledgments
I would like to thank  A.~Uhlmann  for valuable remarks.

\end{document}